\documentclass[prl,twocolumn,longbibliography,superscriptaddress]{revtex4-1}
\usepackage{amsmath}
\usepackage{amssymb}
\usepackage{amsthm}
\usepackage{amsfonts}
\usepackage{listings}
\usepackage{enumerate}
\usepackage{latexsym}
\usepackage{psfrag}
\usepackage{bm}
\usepackage[all]{xy}
\usepackage{graphicx}
\usepackage{subfigure}
\usepackage[pdftex,colorlinks=false]{hyperref}
\usepackage{color}
\usepackage{mathtools}
\usepackage{ gensymb }
\usepackage{lipsum}

\usepackage{float}
\begin{document}

\title{Interfacial superconductivity induced by single-quintuple-layer
  Bi$_2$Te$_3$ on top of FeTe forming van-der-Waals heterostructure}

\author{Hailang Qin}
\affiliation{Shenzhen Institute for Quantum Science and Engineering, and Department of Physics, Southern University of Science and Technology, Shenzhen 518055, China}
\affiliation{Shenzhen Key Laboratory of Quantum Science and Engineering, Shenzhen 518055, China}
\author{Bin Guo}
\author{Linjing Wang}
\author{Meng Zhang}
\author{Bochao Xu}
\author{Kaige Shi}
\author{Tianluo Pan}
\author{Liang Zhou}
\affiliation{Shenzhen Institute for Quantum Science and Engineering, and
  Department of Physics, Southern University of Science and Technology, Shenzhen
  518055, China}
\author{Yang Qiu}
\affiliation{Materials Characterization and Preparation Center, Southern University of Science and Technology, Shenzhen 518055, China}
\author{Bin Xi}
\affiliation{College of Physics Science and Technology, Yangzhou University, Yangzhou 225002, China}
\author{Iam Keong Sou}
\affiliation{Department of Physics, the Hong Kong University of Science and Technology, Hong Kong, China}
\author{Dapeng Yu}

\author{Wei-Qiang Chen}
\author{Hongtao He}
\email{heht@sustech.edu.cn}
\author{Fei Ye}
\email{yef@sustech.edu.cn}
\author{Jia-Wei Mei}
\email{meijw@sustech.edu.cn}
\affiliation{Shenzhen Institute for Quantum Science and Engineering, and Department of Physics, Southern University of Science and Technology, Shenzhen 518055, China}

\author{Gan Wang}
\email{wangg@sustech.edu.cn}
\affiliation{Shenzhen Institute for Quantum Science and Engineering, and Department of Physics, Southern University of Science and Technology, Shenzhen 518055, China}
\affiliation{Shenzhen Key Laboratory of Quantum Science and Engineering, Shenzhen 518055, China}

\date{\today}

\begin{abstract}
  We report the first clear observation of interfacial superconductivity
  on top of FeTe(FT) covered by one quintuple-layer Bi$_2$Te$_3$(BT)
  forming van-der-Waals heterojunction. Both transport and scanning
  tunneling spectroscopy measurements confirm the occurrence of
  superconductivity at a transition temperature T$_c$ = 13~K, when a
  single-quintuple-layer BT is deposited on the non-superconducting FT
  surface.  The superconductivity gap decays exponentially with the
  thickness of BT, suggesting it occurs at the BT-FT interface and the
  proximity length is above 5 - 6~nm.  We also measure the
  work function's dependence on the thickness of BT, implying a
  charge transfer may occur at the BT/FT interface to introduce hole
  doping into the FT layer, which may serve as a possible candidate for
  the superconducting mechanism. Our BT/FT heterojunction provides a
  clean system to study the unconventional interfacial
  superconductivity.
\end{abstract}

\maketitle

The interfacial superconductivity has been of great interest
~\cite{Ginzburg1964,Miller1973,Ahn1999,Ahn2003,Reyren2007,Gozar2008,Pereiro2011},
since it may help to resolve the mystery of high temperature
superconductors with layered structures and to search for the topological superconductors~\cite{Wang2014,He2014,Zhang2018}.  There are several mechanisms
proposed for the interfacial superconductivity, including edge misfit
dislocation defect~\cite{Miller1973,Fogel1996}, electric field
gating~\cite{Ahn1999,Ahn2003} as well as chemical doping (e.g.  charge
transfer) ~\cite{Reyren2007,Gozar2008}. In order to pin down the
superconducting mechanism unambiguously, the high quality sample with
atomic abruptness at the interfaces is
required. In the past decade, van der Waals (vdW) epitaxy turns out to be an
effective way to grow two-dimensional interfacial
superconductors~\cite{Xu2014,Wang2014,He2014,Zhang2018}, where the
interface is usually of high quality even against the large lattice
mismatch since  the
interfacial interaction at the heterojunctions is of vdW type\cite{Koma1992}. 

Bi$_2$Te$_3$/Fe$_{1+x}$Te is one of the first realized van der Waals
heterostructures which hosts an interfacial superconductivity between
two non-superconductors Bi$_2$Te$_3$ and
Fe$_{1+x}$Te~\cite{Wang2014,He2014}, where Fe$_{1+x}$Te may have an
unusual mechanism for superconductivity and magnetic order of '11'
group iron based superconductors~\cite{Bao2009, Xia2009}. A
transport study showed that an optimal superconductivity (T$_c$ = 11.5 K)
only can be developed when the Bi$_2$Te$_3$ layer reaches a thickness
of 5 quintuple-layer (QL) while single QL Bi$_2$Te$_3$/Fe$_{1+x}$Te only
possesses a low T$_{c}$ around 2 K, indicating Bi$_2$Te$_3$ thickness may
be crucial for the formation of superconductivity~\cite{He2014}. In the past years, great efforts have
been devoted to studying the intriguing superconductivity of
Bi$_2$Te$_3$/Fe$_{1+x}$Te~\cite{Du2015, kunchur2015, manna2017,
  fabian2018, Raj2018}. For instance, Gu et al. reported 
a scanning tunneling microscopy (STM) study on the Bi$_2$Te$_3$ grown on Fe$_{1+x}$Te, however, the energy gap of superconductivity didn't appear even Bi$_2$Te$_3$ is thicker than 6 QLs, instead
a merging of Dirac electrons and the correlation effect was revealed in the scanning tunneling spectroscopy (STS), implying the thickness of Bi$_2$Te$_3$ is not the exclusive 
cause of superconductivity~\cite{Du2015}. In 2017, Manna et al. studied a reverse structure, FeTe grown on bulk Bi$_2$Te$_3$ crystal,  and probed an energy gap of superconductivity co-existing with bi-collinear antiferromagnetic (AFM) order~\cite{manna2017}, however, the gap revealed T$_c$ is lower than 6 K, inconsistent with the previous transport transport results~\cite{He2014}, as in their sample the Bi$_2$Te$_3$ layer is thicker than the optimal 5 QLs. Obviously, the mechanism for Bi$_2$Te$_3$/Fe$_{1+x}$Te superconductivity is still a matter of debate, including the location of superconductivity and the possible driving factors, nevertheless, the revealed correlation effect and unusual AFM order made the system very attractive for studying the unusual mechanism for iron based superconductors. 

In this work, we report the STM study of high quality Bi$_2$Te$_3$/FeTe interface grown on SrTiO$_3$(001) (STO) substates. Remarkably, we found that an interfacial superconductivity immediately appeared once one QL Bi$_2$Te$_3$ patched the FeTe surface, with a high T$_c$ around 13 K confirmed by STS and transport study simultaneously. Furthermore, we present the first superconducting energy gap evolution of Bi$_2$Te$_3$/FeTe interface, implying that the superconductivity on the thicker Bi$_2$Te$_3$(2-5QL) layer is proximity-induced and the superconductivity is essentially located at the Bi$_2$Te$_3$-FeTe interface (including the cases of within one QL of Bi$_2$Te$_3$ and within one monolayer of FeTe). Via a deliberate measurement of work function change with Bi$_2$Te$_3$ thickness, we also found that the superconductivity is likely induced by a hole-doping effect on FeTe due to the band mis-alignment between Bi$_2$Te$_3$ and FeTe. We believe that our findings will contribute the key facts for resolving the unconventional interfacial superconductivity in the Bi$_2$Te$_3$/FeTe interface.

\begin{figure}[t]
\centering
\includegraphics[width=0.85\columnwidth]{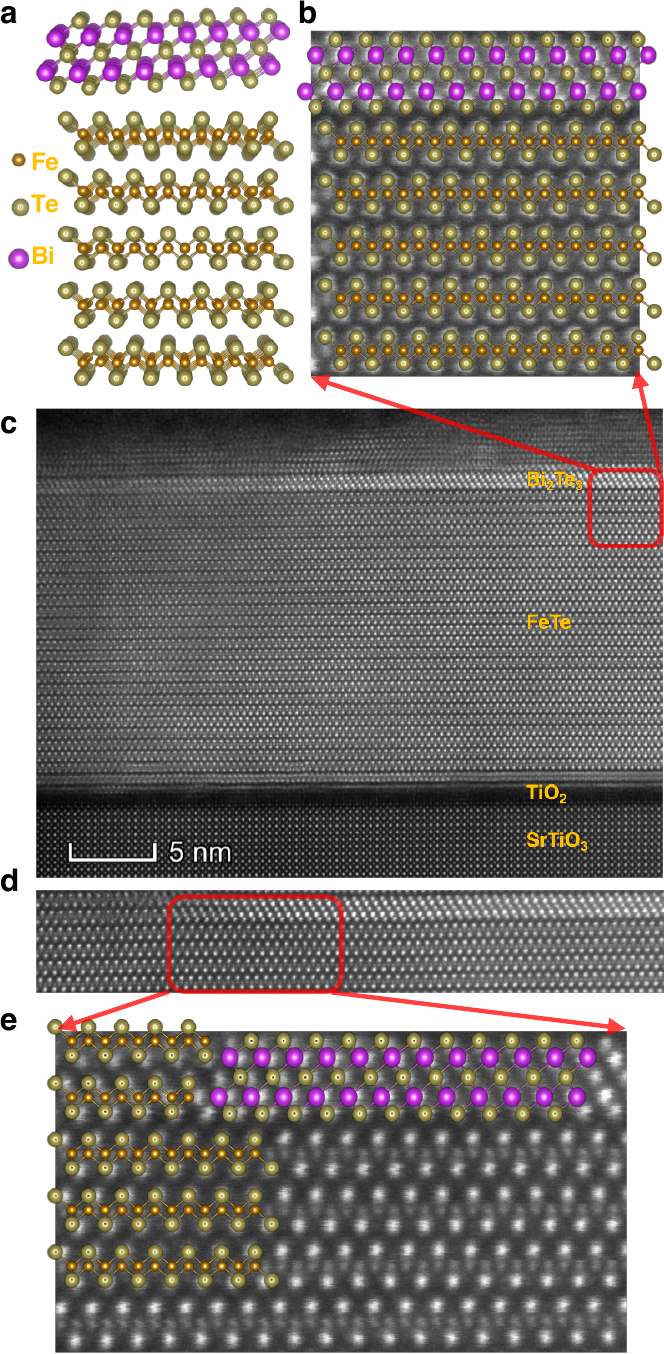}
\caption{(a) A schematic illustration for 1 QL Bi$_2$Te$_3$ grown on FeTe. (b) A lattice comparison for a Bi$_2$Te$_3$/FeTe interface. (c) A HRSTEM image containing a Bi$_2$Te$_3$(1 QL), a FeTe(20nm) thin film and a STO substrate simultaneously. (d) A HRTEM image of a lateral heterojunction formed by a Bi$_2$Te$_3$ quintuple layer and a FeTe atomic step.(e) A lattice comparison for the lateral heterojunction and its nearby FeTe}
\label{fig:figure1}
\end{figure}

The samples were grown on single crystal 0.7\%~wt Nb-doped STO(001)
substrates. The STO substrate was first gradually heated to about 950\celsius~ by
electron-beam heating and annealed at this temperature for about 30 minutes. The
substrate was then transferred \textit{in-situ} to an adjacent Createc molecular beam
epitaxy (MBE) system for sample growth. The FeTe thin films were grown in the
MBE chamber by co-evaporation of high purity Fe (99.995\%) and Te (99.999\%)
onto the substrate held at around 300~\celsius, with a Te rich flux ratio to ensure the stoichiometry. The thickness of FeTe layer is around 20~nm. 
The Bi$_2$Te$_3$ thin films were then
grown by evaporating Bi$_2$Te$_3$ compound source at the substrate temperature of
about 225~\celsius, with a growth rate around 2.5 \AA/min for ensuring the precise QL control. Reflective high energy electron diffraction (RHEED) patterns (supplementary 1) kept streaky during the entire growth, indicating a layer-by-layer growth mode dominating the epitaxy.  The samples were then transferred \textit{in-situ} to a SPECS
Joule-Thomson (JT) STM system. The base pressure in the MBE chamber and STO
chamber was about $3\times10^{-10}$ mbar and $5\times10^{-10}$ mbar,
respectively. The STM measurements were performed at 1.1 K (unless otherwise
specified) with etched tungsten tips, which were sputtered with an Ar$^{+}$ ion
sputter gun and tested on a reference Au(111). A high resolution scanning transmission electron
microscopy(HRSTEM) was employed for a cross-sectional characterization on a
Bi$_2$Te$_3$(1QL)/FeTe(20nm) bilayer sample cut by a focus ion beam technique.
Transport measurements were carried out with a standard four-point probe method using a commercial Quantum
Design physical property measurement (PPMS) system.  The
ultraviolet photoemission spectroscopy (UPS) measure-ments were carried out
using a SPECS PHOIBOS 150 hemispherical energy analyzer and a light source from
a helium discharge lamp (He I, photon energy 21.218 eV).

\begin{figure*}[t]
\centering
\includegraphics[width=2\columnwidth]{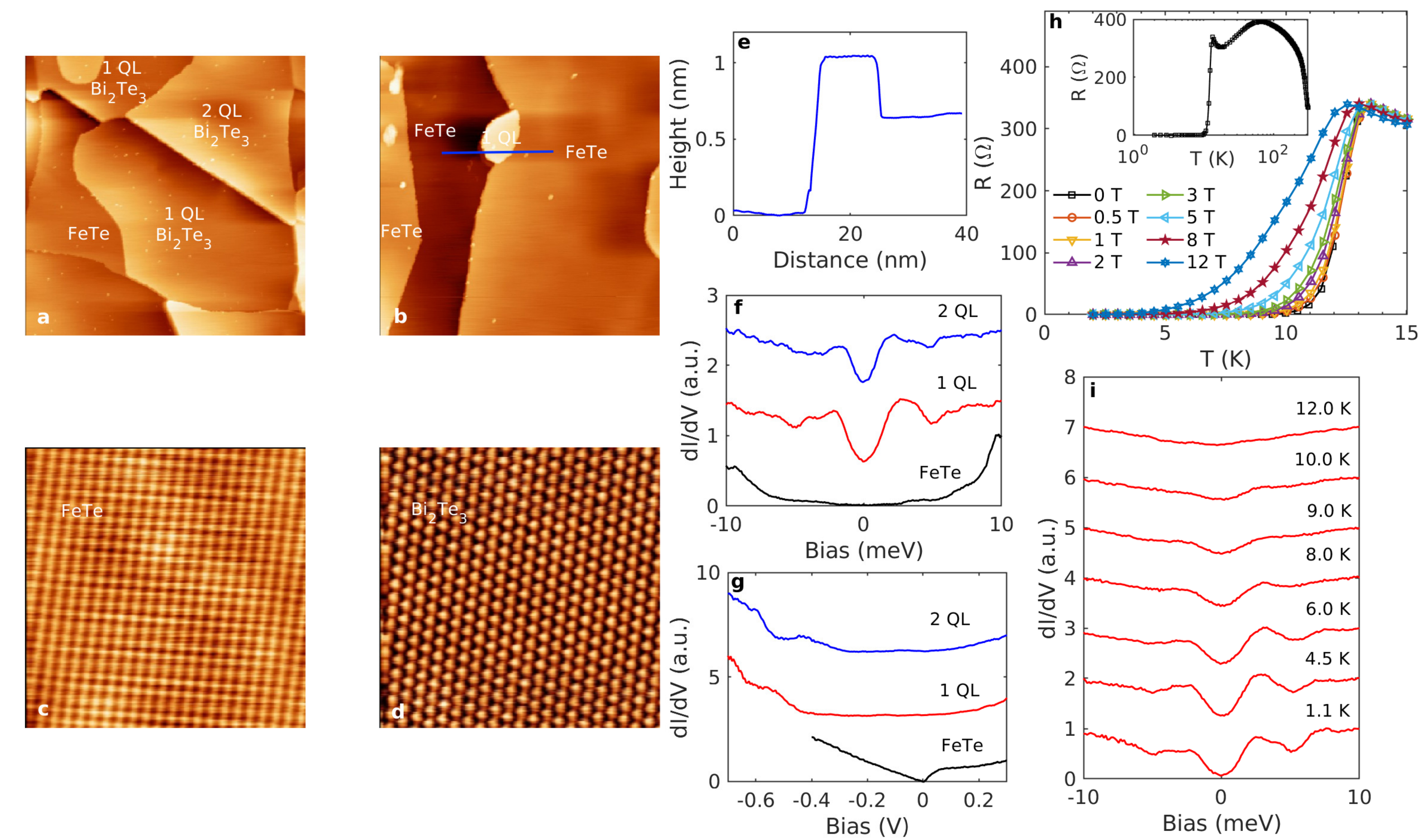}
\caption{(a) A STM topographic image of an area containing exposed FeTe, 1 QL and
  2 QLs of Bi$_2$Te$_3$ (size: $220\times220$~nm$^2$, $V_{\text{Bias}} = 2.0$~V,
  $I_{\text{Tunnel}} = 50$~pA, color scale: 1.57~nm). (b) A STM topographic
  image showing how Bi$_2$Te$_3$ typically grows on FeTe films (size:
  $100\times100$~nm$^2$, $V_{\text{Bias}} = 1.8$~V, $I_{\text{Tunnel}} = 50$~pA,
  color scale:  1.24~nm). (c) An atomic resolution 
  image of the exposed FeTe surfae (size: $8\times8$~nm$^2$, $V_{\text{Bias}} = -10$~mV, $I_{\text{Tunnel}} = 200$~pA,
  color scale: 0.03~nm). (d) An atomic resolution image of the 1 QL Bi$_2$Te$_3$
  surfae (size: $8\times8$~nm$^2$, $V_{\text{Bias}} = 6.5$~mV,
  $I_{\text{Tunnel}} = 400$~pA, color scale: 0.03~nm). (e) A line profile corresponding to the blue line in
  (b). (f). Representative $dI/dV$ spectra aquired on the FeTe surface, 1 QL Bi$_2$Te$_3$, and 2 QL
  Bi$_2$Te$_3$ terraces, respectively (set point: $V_{\text{Bias}} = 10$~mV,
  $I_{\text{Tunnel}} = 50$~pA). (g) Representative $dI/dV$ spectra in a wider
  bias range acquired on the FeTe surface, 1 QL Bi$_2$Te$_3$. and 2 QL
  Bi$_2$Te$_3$ terraces, respectively (set point: $V_{\text{Bias}} = 0.3$~V,
  $I_{\text{Tunnel}} = 50$~pA). (h). Temperature dependent resistance $R$-$T$
  curve of a sample of  1nm Bi$_2$Te$_3$/40 nm FeTe/STO under different magnetic
  fields perpendicular to the interface. The inset in (h) shows the $R$-$T$
  curve of the same sample in a wider temperature range without applying
  magnetic field. (i) Temperature dependent $dI/dV$ spectra acquired
 on the 1 QL Bi$_2$Te$_3$  (set point: $V_{\text{Bias}} = 10 mV$, $I_{\text{Tunnel}} = 50 pA$).The texts in the all the STM images indicate the respective areas identified. Measured temperature is 1.1~K.}
\label{fig:figure2}
\end{figure*}
  Fig.~\ref{fig:figure1}~(a) suggests a cross-sectional schematic illustration of 1 QL Bi$_2$Te$_3$ grown on FeTe, well matching the lattice of the red square indicated region(Fig.~\ref{fig:figure1}~(b)) shown in the HRSTEM Fig. ~\ref{fig:figure1}~(c), evidencing an sharp vdW interface without lattice distortion or inter-diffusion. Furthermore, a typical lateral B$_2$Te$_3$-FeTe heterojunction near a FeTe atomic step is imaged in Fig. ~\ref{fig:figure1}~(d) and~\ref{fig:figure1}~(e),  demonstrating a step-flow epitaxy. It is worth mentioning that no interstitial Fe atoms were detected by a careful examining the FeTe lattice shown in Fig. ~\ref{fig:figure1}~(c). Combining with the 1:1 ratio of Fe/Te revealed by chemical analysis(see supplementary materials), we assert that the FeTe film is of strict stoichiometry, similar as previous reported stoichiometric FeTe thin films grown by MBE with Te rich flux~\cite{Li2016}.

The STM studies were carried out on more than 20 samples and the results are repeatable and consistent. 
Fig. ~\ref{fig:figure2}~(a) is a typical topographic image after growth of
about 1$\sim$2~QL of Bi$_2$Te$_3$, inside which we acquire  atomic-resolution images of exposed FeTe surface and 1~QL Bi$_2$Te$_3$ surface shown in Fig.~\ref{fig:figure2}~(c) and (d), respectively, representing a perfect lattice without any adatoms and vacancies being observed. The growth time for the sample in Fig. 2(b) was extremely short to observe the growth behavior, showing a nucleation of Bi$_2$Te$_3$ takes place on the
atomic step of FeTe, consistent with cross-sectional HRSTEM results shown in Fig.~\ref{fig:figure1}. 
Fig.~\ref{fig:figure2}~(e) is the line profile corresponding to the blue line in Fig.~\ref{fig:figure2}~(b). The step height of about
0.6 nm and 1.0 nm correspond to one unit cell of FeTe and 1~QL layer of Bi$_2$Te$_3$, respectively. Representative $dI/dV$ spectra in a narrow bias range acquired on the FeTe, 1~QL
Bi$_2$Te$_3$, and 2~QL Bi$_2$Te$_3$ terraces are shown in
Fig.~\ref{fig:figure2}~(f). $dI/dV$ measures the local DOS near the Fermi level. The spectra acquired
on the 1~QL Bi$_2$Te$_3$ terrace show two clear coherent superconducting peaks around  2.5~meV. This is in clear contrast to the spectrum acquired on the FeTe surface, which shows no coherent peaks, but a valley shape spectrum with a flat bottom. On the 2~QL Bi$_2$Te$_3$, the spectrum also exhibits two coherent superconducting
peaks; however, the peaks are smaller at around 1.8~meV.
The spectra on each terrace also vary
little, indicating a good uniformity of the superconductivity. The temperature dependent $dI/dV$ spectra on the 1~QL Bi$_2$Te$_3$ surface are
shown in Fig.~\ref{fig:figure2}~(i), while the results on the 2~QL are in the
Supplementary Materials~\cite{Note1}.  While no significant change in the
intensity of the coherence peak for the spectra is observed for temperature below 6~K, there
is a significant decrease in the peak intensity at temperatures above 6~K. The
gap magnitude is discernible up to about 10~K. At 12~K, there is no clear gap
observed, consistent with followed transport revealed T$_c$. 
Typical $dI/dV$ spectra in a wider bias range obtained in the FeTe, 1 QL Bi$_2$Te$_3$, and 2 QL Bi$_2$Te$_3$
surfaces are plotted in Fig.~\ref{fig:figure2}~(g). According to the dI/dV spectrum Bi$_2$Te$_3$ film of 1~QL, there is discernibly weak and nearly constant density of states from 0~eV (Fermi
level) to -0.43~eV. Below about -0.43~eV, the density of states start to rise rapidly with a kink at about -0.55~eV. On the Bi$_2$Te$_3$ film of 2 QL, there are two kinks around 0.08~eV and -0.44~eV, likely corresponding to the conduction band
minimum (CBM) and valence band maximum (VBM), respectively. Comparing with the
$dI/dV$ spectrum on the thick Bi$_2$Te$_3$ film~\cite{Note1}, it indicates that the 1~QL Bi$_2$Te$_3$ film
and 2~QL Bi$_2$Te$_3$ film are both more n-type doped. While this is qualitatively
consistent with what is observed by Xu et al.~\cite{Xu2015}, the  Fermi level shifts on the 1~QL and 2~QL Bi$_2$Te$_3$ are more significant, indicating the electron doping effect from the FeTe film likely plays an important role.

The temperature dependent in-plane resistance ($R$-$T$ curve) of a
Bi$_2$Te$_3$/FeTe/STO device with an average Bi$_2$Te$_3$ thickness of around
1~QL is shown in Fig.~\ref{fig:figure2}~(h). In the zero field (the inset in
Fig.~\ref{fig:figure2}~(h)), a broad transition at about 65~K is observed, likely related to the magnetic and/or structural transition of FeTe films~\cite{Bao2009,Li2009,Roesler2011,Rodriguez2011,Mizuguchi2012,Zaliznyak2012,Koz2013}. The
$R$–$T$ curve shows a superconducting transition with an onset temperature of $T^{\text{onset}}_c=13$~K
and zero resistance temperature of $T^0_c=7.5$~K. Before $T_c^{\text{onset}}$, there is a
upturn in $R$-$T$ curve with decreasing temperature. The magnetic field (applied perpendicular to the
sample interface) suppresses the superconductivity, and 12~T field is far away
to completely kill the superconductivity, indicating a very high critical field, which is a characteristic feature of FeTe based superconductors~\cite{Si2010}.

\begin{figure}[t]
\centering
\includegraphics[width=\columnwidth]{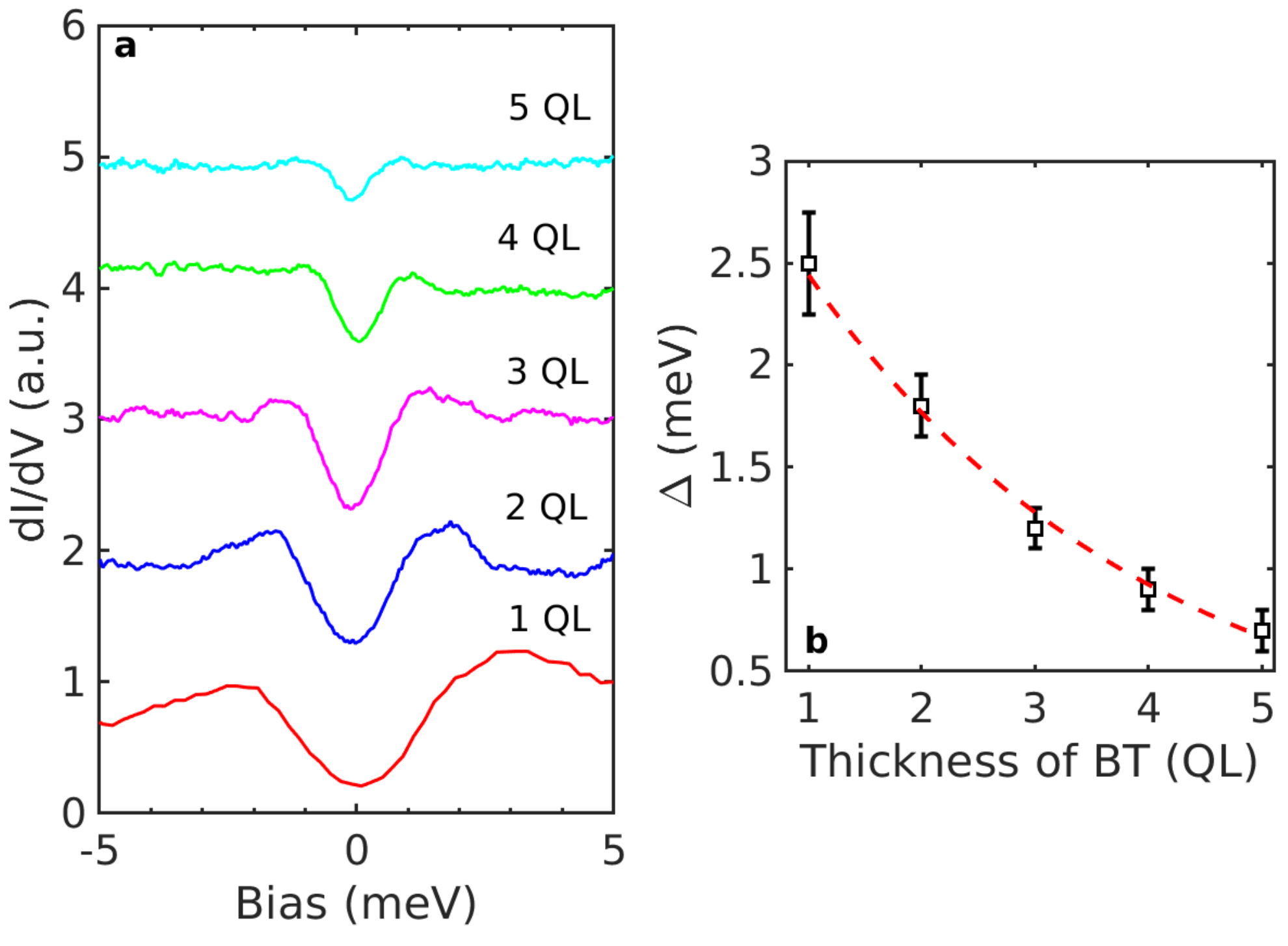}
\caption{(a) Representative dI/dV spectra acquired on 1 QL, 2 QL, 3 QL, 4 QL, and 5 QL Bi$_2$Te$_3$ surfaces, respectively. Spectra were shifted vertically for clarity. (b) Gap magnitude versus the Bi$_2$Te$_3$ thickness.
}
\label{fig:figure3}
\end{figure}
The $dI/dV$ spectra were acquired on samples with
thicker Bi$_2$Te$_3$ films up to about 5-6 QL and are all shown in Fig.~\ref{fig:figure3}~(a). As can
be seen, all of these spectra exhibit two coherent peaks, whereas the gap
magnitude decreases with increasing Bi$_2$Te$_3$ thickness. The gap magnitude $\Delta$
of the spectra is plotted in Fig.~\ref{fig:figure3}~(b) with respect to the number of QLs of Bi$_2$Te$_3$.
The $\Delta$ shows an exponential decay with a relationship of $\Delta=3.376e^{-\frac{N}{3.087}}$ (in unit of meV, N is
the number of QLs of Bi$_2$Te$_3$) as obtained from the fitting, consistent with the characteristic of proximity-induced
superconductivity~\cite{Xu2014}, from which we can infer that the superconductivity is not enhanced in any way when the Bi$_2$Te$_3$  is 2 QLs or more than that when the Bi$_2$Te$_3$ is only 1 QL; instead, the superconductivity locates in the 1 QL Bi$_2$Te$_3$/FeTe interface, and becomes
weaker on the Bi$_2$Te$_3$ surface due to the proximity effect.

\begin{figure}[t]
\centering
\includegraphics[width=\columnwidth]{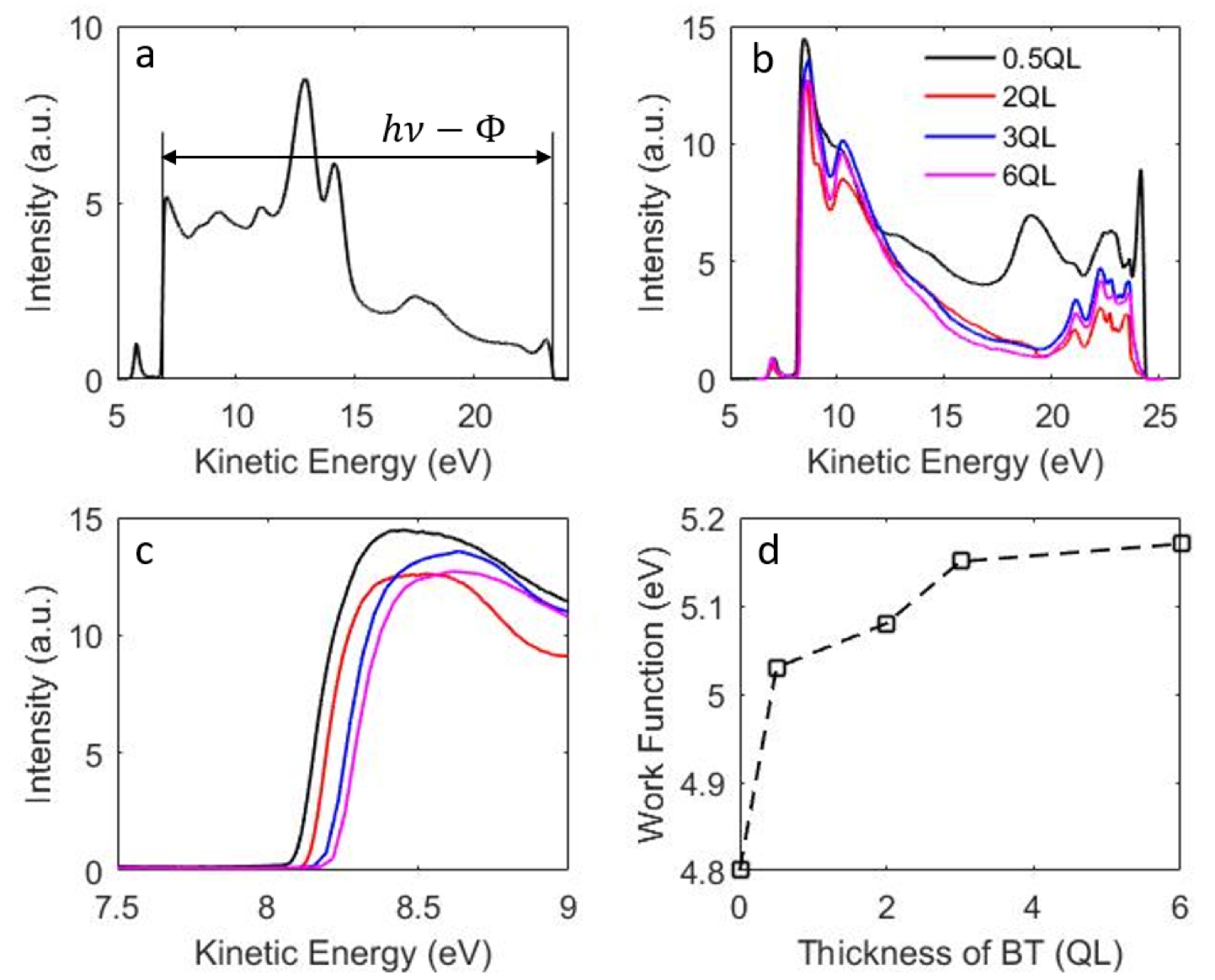}
\caption{(a). UPS spectrum of FeTe film taken with -2 V bias applied to the sample. (b). UPS spectra of Bi$_2$Te$_3$/FeTe samples taken with -3 V bias with average Bi$_2$Te$_3$ thickness of 0.5 QL (black), 2 QL (red), 3 QL (blue), and 6 QL (magenta), respectively. (c). Zoom-in view of (b) around the inelastic cutoff region at lower kinetic energy. (d) Extracted work function from UPS spectra in (a) and (b).
}
\label{fig:figure4}
\end{figure}
To further unveil the charge transfer mechanism between Bi$_2$Te$_3$ and FeTe,
we also \textit{in-situ} measured the work function of the FeTe film before and after the growing of Bi$_2$Te$_3$ film with average thickness of 0.5~QL, 2~QL, 3~QL, and 6~QL , respectively using UPS at 15~K. The UPS result of FeTe film is shown in Fig.~\ref{fig:figure4}~(a). A voltage -2 V was applied between the sample and the spectrometer so that the photoelectrons were accelerated and then the low energy inelastic electrons could be distinguished from secondary electrons generated in the spectrometer by impact ~\cite{Schlaf2013}, which can be seen in  Fig.~\ref{fig:figure4}~(a) as a small peak near 5.5 eV. The work function is calculated with the formula $\Phi=h\nu-\Delta E$, where $h\nu$ is the photon energy and $\Delta E$ is the spectrum width, the distance between the low energy cutoff and the Fermi edge as shown in Fig.~\ref{fig:figure4}~(a). Then we measured the photoemission spectrums of Bi$_2$Te$_3$ covered-~FeTe with increasing thickness of Bi$_2$Te$_3$ and the results are shown in Fig.~\ref{fig:figure4}~(b). 
The zoom-in view of the low kinetic energy region of Fig.~\ref{fig:figure4}~(b) is shown in Fig.~\ref{fig:figure4}~(c), depicting that the inelastic edge offsets to the high kinetic energy region as the thickness of Bi$_2$Te$_3$ being increased, while the fermi edge in the high energy end is pinned in all the spectrums as they are determined by the accelerating voltage(-~3 V) solely. The extracted work functions are shown in Fig.~\ref{fig:figure4}~(d). As can be seen, the work function of FeTe film is about 4.80 eV. With the growth of Bi$_2$Te$_3$, the work function increases gradually, and starts to saturate when the Bi$_2$Te$_3$ thickness reaches about 3 QL. This result clearly suggests that there is a doping effect when the Bi$_2$Te$_3$ film is deposited onto the FeTe film, i.e., FeTe is hole-doped and Bi$_2$Te$_3$ film is electron-doped, qualitatively consistent the STS revealed charge transfer in Fig.~\ref{fig:figure2}~(g).

In the heterostructure of Bi$_2$Te$_3$/FeTe in our study, neither Bi$_2$Te$_3$ or FeTe is superconducting. Upon doping, Bi$_2$Te$_3$ is only superconducting with a maximum reported transition temperature of 5.5~K~\cite{Hor2010}. On the other hand, FeTe is superconducting upon doping with either oxygen~\cite{Si2010} or selenium~\cite{fang2008} with a transition temperature of about 14~K, which is surprisingly close to the transition temperature in our study. Our work function measurements clearly show that there is a hole doping effect on the FeTe layer in the heterostructure. Therefore, it is plausible that the single FeTe layer become superconductivity due to the loss of electrons induced by Bi$_2$Te$_3$ covering, leading to the interface superconductivity observed in our study. While it still remains a challenge on how this doping leads to the observed superconductivity, our results provide more insights for future theoretic work to understand this doping induced change on electronic and magnetic properties of a single FeTe layer, which may also shed light on resolving the mystery of iron based superconductivity.

In conclusion, the superconductivity of Bi$_2$Te$_3$/FeTe system is
studied by using the low temperature scanning tunneling microscopy. It
is found that the superconductivity in the Bi$_2$Te$_3$/FeTe system
can be induced by only one QL of Bi$_2$Te$_3$, which is also confirmed
by the transport measurements. Scanning tunneling spectroscopy further
shows that the superconducting gap decays exponentially with the
increasing Bi$_2$Te$_3$ thickness, implying the superconductivity in
Bi$_2$Te$_3$ bulk is actually proximity-induced by the interface.  Our results
provide unambiguous evidence that the superconductivity in the
Bi$_2$Te$_3$/FeTe system is located around the Bi$_2$Te$_3$/FeTe
interface. As for the superconducting mechanism, it is found that the
doping of FeTe may be possible to generate superconductivity in the Bi$_2$Te$_3$/FeTe vdW hetero-junctions.

\acknowledgements{We thanks H. H. Wen, L. Yu and P. A. Marchetti for discussions. This work was supported by the National Natural Science
Foundation of China (NO. 61734008 and 11774143), the National Key Research and Development Program of China
(No.~2018YFA0307100, No.~2016YFA0301703), the Natural Science Foundation of Guangdong Province
(No.~2015A030313840, No.~2017A030313033), the State Key Laboratory of
Low-Dimensional Quantum Physics (No.~KF201602), Technology and Innovation
Commission of Shenzhen Municipality ( ZDSYS20170303165926217 and JCYJ20170412152334605). J.W.M was partially supported by the program 
for Guangdong Introducing Innovative and Entrepreneurial Teams (No. 2017ZT07C062).}

\bibliography{../FeTe}

\begin{thebibliography}{33}%
\makeatletter
\providecommand \@ifxundefined [1]{%
 \@ifx{#1\undefined}
}%
\providecommand \@ifnum [1]{%
 \ifnum #1\expandafter \@firstoftwo
 \else \expandafter \@secondoftwo
 \fi
}%
\providecommand \@ifx [1]{%
 \ifx #1\expandafter \@firstoftwo
 \else \expandafter \@secondoftwo
 \fi
}%
\providecommand \natexlab [1]{#1}%
\providecommand \enquote  [1]{``#1''}%
\providecommand \bibnamefont  [1]{#1}%
\providecommand \bibfnamefont [1]{#1}%
\providecommand \citenamefont [1]{#1}%
\providecommand \href@noop [0]{\@secondoftwo}%
\providecommand \href [0]{\begingroup \@sanitize@url \@href}%
\providecommand \@href[1]{\@@startlink{#1}\@@href}%
\providecommand \@@href[1]{\endgroup#1\@@endlink}%
\providecommand \@sanitize@url [0]{\catcode `\\12\catcode `\$12\catcode
  `\&12\catcode `\#12\catcode `\^12\catcode `\_12\catcode `\%12\relax}%
\providecommand \@@startlink[1]{}%
\providecommand \@@endlink[0]{}%
\providecommand \url  [0]{\begingroup\@sanitize@url \@url }%
\providecommand \@url [1]{\endgroup\@href {#1}{\urlprefix }}%
\providecommand \urlprefix  [0]{URL }%
\providecommand \Eprint [0]{\href }%
\providecommand \doibase [0]{http://dx.doi.org/}%
\providecommand \selectlanguage [0]{\@gobble}%
\providecommand \bibinfo  [0]{\@secondoftwo}%
\providecommand \bibfield  [0]{\@secondoftwo}%
\providecommand \translation [1]{[#1]}%
\providecommand \BibitemOpen [0]{}%
\providecommand \bibitemStop [0]{}%
\providecommand \bibitemNoStop [0]{.\EOS\space}%
\providecommand \EOS [0]{\spacefactor3000\relax}%
\providecommand \BibitemShut  [1]{\csname bibitem#1\endcsname}%
\let\auto@bib@innerbib\@empty
\bibitem [{\citenamefont {Ginzburg}(1964)}]{Ginzburg1964}%
  \BibitemOpen
  \bibfield  {author} {\bibinfo {author} {\bibfnamefont {V.~L.}\ \bibnamefont
  {Ginzburg}},\ }\bibfield  {title} {\enquote {\bibinfo {title} {On surface
  superconductivity},}\ }\href
  {http://www.sciencedirect.com/science/article/pii/0031916364906729}
  {\bibfield  {journal} {\bibinfo  {journal} {Physics Letters}\ }\textbf
  {\bibinfo {volume} {13}},\ \bibinfo {pages} {101--102} (\bibinfo {year}
  {1964})}\BibitemShut {NoStop}%
\bibitem [{\citenamefont {Miller}\ \emph {et~al.}(1973)\citenamefont {Miller},
  \citenamefont {Strongin}, \citenamefont {Kammerer},\ and\ \citenamefont
  {Streetman}}]{Miller1973}%
  \BibitemOpen
  \bibfield  {author} {\bibinfo {author} {\bibfnamefont {D.~L.}\ \bibnamefont
  {Miller}}, \bibinfo {author} {\bibfnamefont {Myron}\ \bibnamefont
  {Strongin}}, \bibinfo {author} {\bibfnamefont {O.~F.}\ \bibnamefont
  {Kammerer}}, \ and\ \bibinfo {author} {\bibfnamefont {B.~G.}\ \bibnamefont
  {Streetman}},\ }\bibfield  {title} {\enquote {\bibinfo {title}
  {Superconductivity at the surface of pbte},}\ }\href {\doibase
  10.1103/PhysRevB.8.4416} {\bibfield  {journal} {\bibinfo  {journal} {Phys.
  Rev. B}\ }\textbf {\bibinfo {volume} {8}},\ \bibinfo {pages} {4416--4419}
  (\bibinfo {year} {1973})}\BibitemShut {NoStop}%
\bibitem [{\citenamefont {Ahn}\ \emph {et~al.}(1999)\citenamefont {Ahn},
  \citenamefont {Gariglio}, \citenamefont {Paruch}, \citenamefont {Tybell},
  \citenamefont {Antognazza},\ and\ \citenamefont {Triscone}}]{Ahn1999}%
  \BibitemOpen
  \bibfield  {author} {\bibinfo {author} {\bibfnamefont {C.~H.}\ \bibnamefont
  {Ahn}}, \bibinfo {author} {\bibfnamefont {S.}~\bibnamefont {Gariglio}},
  \bibinfo {author} {\bibfnamefont {P.}~\bibnamefont {Paruch}}, \bibinfo
  {author} {\bibfnamefont {T.}~\bibnamefont {Tybell}}, \bibinfo {author}
  {\bibfnamefont {L.}~\bibnamefont {Antognazza}}, \ and\ \bibinfo {author}
  {\bibfnamefont {J.-M.}\ \bibnamefont {Triscone}},\ }\bibfield  {title}
  {\enquote {\bibinfo {title} {Electrostatic modulation of superconductivity in
  ultrathin gdba2cu3o7-x films},}\ }\href {\doibase
  10.1126/science.284.5417.1152} {\bibfield  {journal} {\bibinfo  {journal}
  {Science}\ }\textbf {\bibinfo {volume} {284}},\ \bibinfo {pages} {1152--1155}
  (\bibinfo {year} {1999})}\BibitemShut {NoStop}%
\bibitem [{\citenamefont {Ahn}\ \emph {et~al.}(2003)\citenamefont {Ahn},
  \citenamefont {Triscone},\ and\ \citenamefont {Mannhart}}]{Ahn2003}%
  \BibitemOpen
  \bibfield  {author} {\bibinfo {author} {\bibfnamefont {C.~H.}\ \bibnamefont
  {Ahn}}, \bibinfo {author} {\bibfnamefont {J.-M.}\ \bibnamefont {Triscone}}, \
  and\ \bibinfo {author} {\bibfnamefont {J.}~\bibnamefont {Mannhart}},\
  }\bibfield  {title} {\enquote {\bibinfo {title} {Electric field effect in
  correlated oxide systems},}\ }\href {https://doi.org/10.1038/nature01878}
  {\bibfield  {journal} {\bibinfo  {journal} {Nature}\ }\textbf {\bibinfo
  {volume} {424}},\ \bibinfo {pages} {1015} (\bibinfo {year}
  {2003})}\BibitemShut {NoStop}%
\bibitem [{\citenamefont {Reyren}\ \emph {et~al.}(2007)\citenamefont {Reyren},
  \citenamefont {Thiel}, \citenamefont {Caviglia}, \citenamefont {Kourkoutis},
  \citenamefont {Hammerl}, \citenamefont {Richter}, \citenamefont {Schneider},
  \citenamefont {Kopp}, \citenamefont {R{\"u}etschi}, \citenamefont {Jaccard},
  \citenamefont {Gabay}, \citenamefont {Muller}, \citenamefont {Triscone},\
  and\ \citenamefont {Mannhart}}]{Reyren2007}%
  \BibitemOpen
  \bibfield  {author} {\bibinfo {author} {\bibfnamefont {N.}~\bibnamefont
  {Reyren}}, \bibinfo {author} {\bibfnamefont {S.}~\bibnamefont {Thiel}},
  \bibinfo {author} {\bibfnamefont {A.~D.}\ \bibnamefont {Caviglia}}, \bibinfo
  {author} {\bibfnamefont {L.~Fitting}\ \bibnamefont {Kourkoutis}}, \bibinfo
  {author} {\bibfnamefont {G.}~\bibnamefont {Hammerl}}, \bibinfo {author}
  {\bibfnamefont {C.}~\bibnamefont {Richter}}, \bibinfo {author} {\bibfnamefont
  {C.~W.}\ \bibnamefont {Schneider}}, \bibinfo {author} {\bibfnamefont
  {T.}~\bibnamefont {Kopp}}, \bibinfo {author} {\bibfnamefont {A.-S.}\
  \bibnamefont {R{\"u}etschi}}, \bibinfo {author} {\bibfnamefont
  {D.}~\bibnamefont {Jaccard}}, \bibinfo {author} {\bibfnamefont
  {M.}~\bibnamefont {Gabay}}, \bibinfo {author} {\bibfnamefont {D.~A.}\
  \bibnamefont {Muller}}, \bibinfo {author} {\bibfnamefont {J.-M.}\
  \bibnamefont {Triscone}}, \ and\ \bibinfo {author} {\bibfnamefont
  {J.}~\bibnamefont {Mannhart}},\ }\bibfield  {title} {\enquote {\bibinfo
  {title} {Superconducting interfaces between insulating oxides},}\ }\href
  {\doibase 10.1126/science.1146006} {\bibfield  {journal} {\bibinfo  {journal}
  {Science}\ }\textbf {\bibinfo {volume} {317}},\ \bibinfo {pages} {1196--1199}
  (\bibinfo {year} {2007})}\BibitemShut {NoStop}%
\bibitem [{\citenamefont {Gozar}\ \emph {et~al.}(2008)\citenamefont {Gozar},
  \citenamefont {Logvenov}, \citenamefont {Kourkoutis}, \citenamefont
  {Bollinger}, \citenamefont {Giannuzzi}, \citenamefont {Muller},\ and\
  \citenamefont {Bozovic}}]{Gozar2008}%
  \BibitemOpen
  \bibfield  {author} {\bibinfo {author} {\bibfnamefont {A.}~\bibnamefont
  {Gozar}}, \bibinfo {author} {\bibfnamefont {G.}~\bibnamefont {Logvenov}},
  \bibinfo {author} {\bibfnamefont {L.~Fitting}\ \bibnamefont {Kourkoutis}},
  \bibinfo {author} {\bibfnamefont {A.~T.}\ \bibnamefont {Bollinger}}, \bibinfo
  {author} {\bibfnamefont {L.~A.}\ \bibnamefont {Giannuzzi}}, \bibinfo {author}
  {\bibfnamefont {D.~A.}\ \bibnamefont {Muller}}, \ and\ \bibinfo {author}
  {\bibfnamefont {I.}~\bibnamefont {Bozovic}},\ }\bibfield  {title} {\enquote
  {\bibinfo {title} {High-temperature interface superconductivity between
  metallic and insulating copper oxides},}\ }\href
  {https://doi.org/10.1038/nature07293} {\bibfield  {journal} {\bibinfo
  {journal} {Nature}\ }\textbf {\bibinfo {volume} {455}},\ \bibinfo {pages}
  {782} (\bibinfo {year} {2008})}\BibitemShut {NoStop}%
\bibitem [{\citenamefont {Pereiro}\ \emph {et~al.}(2011)\citenamefont
  {Pereiro}, \citenamefont {Petrovic}, \citenamefont {Panagopoulos},\ and\
  \citenamefont {Bo{\v{z}}ovi{\'c}}}]{Pereiro2011}%
  \BibitemOpen
  \bibfield  {author} {\bibinfo {author} {\bibfnamefont {Juan}\ \bibnamefont
  {Pereiro}}, \bibinfo {author} {\bibfnamefont {Alexander}\ \bibnamefont
  {Petrovic}}, \bibinfo {author} {\bibfnamefont {Christos}\ \bibnamefont
  {Panagopoulos}}, \ and\ \bibinfo {author} {\bibfnamefont {Ivan}\ \bibnamefont
  {Bo{\v{z}}ovi{\'c}}},\ }\bibfield  {title} {\enquote {\bibinfo {title}
  {Interface superconductivity: History, development and prospects},}\
  }\href@noop {} {\bibfield  {journal} {\bibinfo  {journal} {arXiv preprint
  arXiv:1111.4194}\ } (\bibinfo {year} {2011})}\BibitemShut {NoStop}%
\bibitem [{\citenamefont {Wang}\ \emph {et~al.}(2014)\citenamefont {Wang},
  \citenamefont {He}, \citenamefont {He}, \citenamefont {Liu}, \citenamefont
  {He}, \citenamefont {Wang}, \citenamefont {Lortz}, \citenamefont {Wong},\
  and\ \citenamefont {Sou}}]{Wang2014}%
  \BibitemOpen
  \bibfield  {author} {\bibinfo {author} {\bibfnamefont {Gan}\ \bibnamefont
  {Wang}}, \bibinfo {author} {\bibfnamefont {Qing~Lin}\ \bibnamefont {He}},
  \bibinfo {author} {\bibfnamefont {Hong-Tao}\ \bibnamefont {He}}, \bibinfo
  {author} {\bibfnamefont {Hong-Chao}\ \bibnamefont {Liu}}, \bibinfo {author}
  {\bibfnamefont {Mingquan}\ \bibnamefont {He}}, \bibinfo {author}
  {\bibfnamefont {Jian-Nong}\ \bibnamefont {Wang}}, \bibinfo {author}
  {\bibfnamefont {Rolf}\ \bibnamefont {Lortz}}, \bibinfo {author}
  {\bibfnamefont {George Ke~Lun}\ \bibnamefont {Wong}}, \ and\ \bibinfo
  {author} {\bibfnamefont {Iam~Keong}\ \bibnamefont {Sou}},\ }\bibfield
  {title} {\enquote {\bibinfo {title} {Formation mechanism of superconducting
  fe1+xte/bi2te3 bilayer synthesized via interfacial chemical reactions},}\
  }\href {\doibase 10.1021/cg500292z} {\bibfield  {journal} {\bibinfo
  {journal} {Crystal Growth \& Design}\ }\textbf {\bibinfo {volume} {14}},\
  \bibinfo {pages} {3370--3374} (\bibinfo {year} {2014})}\BibitemShut {NoStop}%
\bibitem [{\citenamefont {He}\ \emph {et~al.}(2014)\citenamefont {He},
  \citenamefont {Liu}, \citenamefont {He}, \citenamefont {Lai}, \citenamefont
  {He}, \citenamefont {Wang}, \citenamefont {Law}, \citenamefont {Lortz},
  \citenamefont {Wang},\ and\ \citenamefont {Sou}}]{He2014}%
  \BibitemOpen
  \bibfield  {author} {\bibinfo {author} {\bibfnamefont {Qing~Lin}\
  \bibnamefont {He}}, \bibinfo {author} {\bibfnamefont {Hongchao}\ \bibnamefont
  {Liu}}, \bibinfo {author} {\bibfnamefont {Mingquan}\ \bibnamefont {He}},
  \bibinfo {author} {\bibfnamefont {Ying~Hoi}\ \bibnamefont {Lai}}, \bibinfo
  {author} {\bibfnamefont {Hongtao}\ \bibnamefont {He}}, \bibinfo {author}
  {\bibfnamefont {Gan}\ \bibnamefont {Wang}}, \bibinfo {author} {\bibfnamefont
  {Kam~Tuen}\ \bibnamefont {Law}}, \bibinfo {author} {\bibfnamefont {Rolf}\
  \bibnamefont {Lortz}}, \bibinfo {author} {\bibfnamefont {Jiannong}\
  \bibnamefont {Wang}}, \ and\ \bibinfo {author} {\bibfnamefont {Iam~Keong}\
  \bibnamefont {Sou}},\ }\bibfield  {title} {\enquote {\bibinfo {title}
  {Two-dimensional superconductivity at the interface of a {Bi$_2$Te$_3$/FeTe}
  heterostructure},}\ }\href {http://dx.doi.org/10.1038/ncomms5247} {\bibfield
  {journal} {\bibinfo  {journal} {Nature Communications}\ }\textbf {\bibinfo
  {volume} {5}},\ \bibinfo {pages} {4247} (\bibinfo {year} {2014})}\BibitemShut
  {NoStop}%
\bibitem [{\citenamefont {Zhang}\ \emph {et~al.}(2018)\citenamefont {Zhang},
  \citenamefont {Fan}, \citenamefont {Wang}, \citenamefont {Zhang},
  \citenamefont {Wang}, \citenamefont {Li}, \citenamefont {He}, \citenamefont
  {Song}, \citenamefont {Ma},\ and\ \citenamefont {Xue}}]{Zhang2018}%
  \BibitemOpen
  \bibfield  {author} {\bibinfo {author} {\bibfnamefont {Yi-Min}\ \bibnamefont
  {Zhang}}, \bibinfo {author} {\bibfnamefont {Jia-Qi}\ \bibnamefont {Fan}},
  \bibinfo {author} {\bibfnamefont {Wen-Lin}\ \bibnamefont {Wang}}, \bibinfo
  {author} {\bibfnamefont {Ding}\ \bibnamefont {Zhang}}, \bibinfo {author}
  {\bibfnamefont {Lili}\ \bibnamefont {Wang}}, \bibinfo {author} {\bibfnamefont
  {Wei}\ \bibnamefont {Li}}, \bibinfo {author} {\bibfnamefont {Ke}~\bibnamefont
  {He}}, \bibinfo {author} {\bibfnamefont {Can-Li}\ \bibnamefont {Song}},
  \bibinfo {author} {\bibfnamefont {Xu-Cun}\ \bibnamefont {Ma}}, \ and\
  \bibinfo {author} {\bibfnamefont {Qi-Kun}\ \bibnamefont {Xue}},\ }\bibfield
  {title} {\enquote {\bibinfo {title} {Observation of interface
  superconductivity in a ${\mathrm{snse}}_{2}$/epitaxial graphene van der waals
  heterostructure},}\ }\href {\doibase 10.1103/PhysRevB.98.220508} {\bibfield
  {journal} {\bibinfo  {journal} {Phys. Rev. B}\ }\textbf {\bibinfo {volume}
  {98}},\ \bibinfo {pages} {220508} (\bibinfo {year} {2018})}\BibitemShut
  {NoStop}%
\bibitem [{\citenamefont {Fogel}\ \emph {et~al.}(1996)\citenamefont {Fogel},
  \citenamefont {Cherkasova}, \citenamefont {Pokhila}, \citenamefont
  {Sipatov},\ and\ \citenamefont {Fedorenko}}]{Fogel1996}%
  \BibitemOpen
  \bibfield  {author} {\bibinfo {author} {\bibfnamefont {Nina~Ya.}\
  \bibnamefont {Fogel}}, \bibinfo {author} {\bibfnamefont {V.~G.}\ \bibnamefont
  {Cherkasova}}, \bibinfo {author} {\bibfnamefont {A.~S.}\ \bibnamefont
  {Pokhila}}, \bibinfo {author} {\bibfnamefont {A.~Yu.}\ \bibnamefont
  {Sipatov}}, \ and\ \bibinfo {author} {\bibfnamefont {A.~I.}\ \bibnamefont
  {Fedorenko}},\ }\bibfield  {title} {\enquote {\bibinfo {title}
  {Superconductivity in the novel semiconducting superlattices},}\ }\href
  {\doibase 10.1007/BF02583671} {\bibfield  {journal} {\bibinfo  {journal}
  {Czechoslovak Journal of Physics}\ }\textbf {\bibinfo {volume} {46}},\
  \bibinfo {pages} {727--728} (\bibinfo {year} {1996})}\BibitemShut {NoStop}%
\bibitem [{\citenamefont {Xu}\ \emph {et~al.}(2014)\citenamefont {Xu},
  \citenamefont {Liu}, \citenamefont {Wang}, \citenamefont {Ge}, \citenamefont
  {Liu}, \citenamefont {Yang}, \citenamefont {Chen}, \citenamefont {Liu},
  \citenamefont {Xu}, \citenamefont {Gao}, \citenamefont {Qian}, \citenamefont
  {Zhang},\ and\ \citenamefont {Jia}}]{Xu2014}%
  \BibitemOpen
  \bibfield  {author} {\bibinfo {author} {\bibfnamefont {Jin-Peng}\
  \bibnamefont {Xu}}, \bibinfo {author} {\bibfnamefont {Canhua}\ \bibnamefont
  {Liu}}, \bibinfo {author} {\bibfnamefont {Mei-Xiao}\ \bibnamefont {Wang}},
  \bibinfo {author} {\bibfnamefont {Jianfeng}\ \bibnamefont {Ge}}, \bibinfo
  {author} {\bibfnamefont {Zhi-Long}\ \bibnamefont {Liu}}, \bibinfo {author}
  {\bibfnamefont {Xiaojun}\ \bibnamefont {Yang}}, \bibinfo {author}
  {\bibfnamefont {Yan}\ \bibnamefont {Chen}}, \bibinfo {author} {\bibfnamefont
  {Ying}\ \bibnamefont {Liu}}, \bibinfo {author} {\bibfnamefont {Zhu-An}\
  \bibnamefont {Xu}}, \bibinfo {author} {\bibfnamefont {Chun-Lei}\ \bibnamefont
  {Gao}}, \bibinfo {author} {\bibfnamefont {Dong}\ \bibnamefont {Qian}},
  \bibinfo {author} {\bibfnamefont {Fu-Chun}\ \bibnamefont {Zhang}}, \ and\
  \bibinfo {author} {\bibfnamefont {Jin-Feng}\ \bibnamefont {Jia}},\ }\bibfield
   {title} {\enquote {\bibinfo {title} {Artificial topological superconductor
  by the proximity effect},}\ }\href {\doibase 10.1103/PhysRevLett.112.217001}
  {\bibfield  {journal} {\bibinfo  {journal} {Phys. Rev. Lett.}\ }\textbf
  {\bibinfo {volume} {112}},\ \bibinfo {pages} {217001} (\bibinfo {year}
  {2014})}\BibitemShut {NoStop}%
\bibitem [{\citenamefont {Koma}(1992)}]{Koma1992}%
  \BibitemOpen
  \bibfield  {author} {\bibinfo {author} {\bibfnamefont {Atsushi}\ \bibnamefont
  {Koma}},\ }\bibfield  {title} {\enquote {\bibinfo {title} {Van der waals
  epitaxy—a new epitaxial growth method for a highly lattice-mismatched
  system},}\ }\href {\doibase https://doi.org/10.1016/0040-6090(92)90872-9}
  {\bibfield  {journal} {\bibinfo  {journal} {Thin Solid Films}\ }\textbf
  {\bibinfo {volume} {216}},\ \bibinfo {pages} {72 -- 76} (\bibinfo {year}
  {1992})}\BibitemShut {NoStop}%
\bibitem [{\citenamefont {Bao}\ \emph {et~al.}(2009)\citenamefont {Bao},
  \citenamefont {Qiu}, \citenamefont {Huang}, \citenamefont {Green},
  \citenamefont {Zajdel}, \citenamefont {Fitzsimmons}, \citenamefont
  {Zhernenkov}, \citenamefont {Chang}, \citenamefont {Fang}, \citenamefont
  {Qian}, \citenamefont {Vehstedt}, \citenamefont {Yang}, \citenamefont {Pham},
  \citenamefont {Spinu},\ and\ \citenamefont {Mao}}]{Bao2009}%
  \BibitemOpen
  \bibfield  {author} {\bibinfo {author} {\bibfnamefont {Wei}\ \bibnamefont
  {Bao}}, \bibinfo {author} {\bibfnamefont {Y.}~\bibnamefont {Qiu}}, \bibinfo
  {author} {\bibfnamefont {Q.}~\bibnamefont {Huang}}, \bibinfo {author}
  {\bibfnamefont {M.~A.}\ \bibnamefont {Green}}, \bibinfo {author}
  {\bibfnamefont {P.}~\bibnamefont {Zajdel}}, \bibinfo {author} {\bibfnamefont
  {M.~R.}\ \bibnamefont {Fitzsimmons}}, \bibinfo {author} {\bibfnamefont
  {M.}~\bibnamefont {Zhernenkov}}, \bibinfo {author} {\bibfnamefont
  {S.}~\bibnamefont {Chang}}, \bibinfo {author} {\bibfnamefont {Minghu}\
  \bibnamefont {Fang}}, \bibinfo {author} {\bibfnamefont {B.}~\bibnamefont
  {Qian}}, \bibinfo {author} {\bibfnamefont {E.~K.}\ \bibnamefont {Vehstedt}},
  \bibinfo {author} {\bibfnamefont {Jinhu}\ \bibnamefont {Yang}}, \bibinfo
  {author} {\bibfnamefont {H.~M.}\ \bibnamefont {Pham}}, \bibinfo {author}
  {\bibfnamefont {L.}~\bibnamefont {Spinu}}, \ and\ \bibinfo {author}
  {\bibfnamefont {Z.~Q.}\ \bibnamefont {Mao}},\ }\bibfield  {title} {\enquote
  {\bibinfo {title} {Tunable ($\ensuremath{\delta}\ensuremath{\pi}$,
  $\ensuremath{\delta}\ensuremath{\pi}$)-type antiferromagnetic order in
  $\ensuremath{\alpha}$-fe(te,se) superconductors},}\ }\href {\doibase
  10.1103/PhysRevLett.102.247001} {\bibfield  {journal} {\bibinfo  {journal}
  {Phys. Rev. Lett.}\ }\textbf {\bibinfo {volume} {102}},\ \bibinfo {pages}
  {247001} (\bibinfo {year} {2009})}\BibitemShut {NoStop}%
\bibitem [{\citenamefont {Xia}\ \emph {et~al.}(2009)\citenamefont {Xia},
  \citenamefont {Qian}, \citenamefont {Wray}, \citenamefont {Hsieh},
  \citenamefont {Chen}, \citenamefont {Luo}, \citenamefont {Wang},\ and\
  \citenamefont {Hasan}}]{Xia2009}%
  \BibitemOpen
  \bibfield  {author} {\bibinfo {author} {\bibfnamefont {Y.}~\bibnamefont
  {Xia}}, \bibinfo {author} {\bibfnamefont {D.}~\bibnamefont {Qian}}, \bibinfo
  {author} {\bibfnamefont {L.}~\bibnamefont {Wray}}, \bibinfo {author}
  {\bibfnamefont {D.}~\bibnamefont {Hsieh}}, \bibinfo {author} {\bibfnamefont
  {G.~F.}\ \bibnamefont {Chen}}, \bibinfo {author} {\bibfnamefont {J.~L.}\
  \bibnamefont {Luo}}, \bibinfo {author} {\bibfnamefont {N.~L.}\ \bibnamefont
  {Wang}}, \ and\ \bibinfo {author} {\bibfnamefont {M.~Z.}\ \bibnamefont
  {Hasan}},\ }\bibfield  {title} {\enquote {\bibinfo {title} {Fermi surface
  topology and low-lying quasiparticle dynamics of parent
  ${\mathrm{fe}}_{1+x}\mathrm{Te}/\mathrm{Se}$ superconductor},}\ }\href
  {\doibase 10.1103/PhysRevLett.103.037002} {\bibfield  {journal} {\bibinfo
  {journal} {Phys. Rev. Lett.}\ }\textbf {\bibinfo {volume} {103}},\ \bibinfo
  {pages} {037002} (\bibinfo {year} {2009})}\BibitemShut {NoStop}%
\bibitem [{\citenamefont {Du}\ \emph {et~al.}(2015)\citenamefont {Du},
  \citenamefont {Du}, \citenamefont {Yang}, \citenamefont {Wang}, \citenamefont
  {Fang}, \citenamefont {Yang},\ and\ \citenamefont {Wen}}]{Du2015}%
  \BibitemOpen
  \bibfield  {author} {\bibinfo {author} {\bibfnamefont {Guan}\ \bibnamefont
  {Du}}, \bibinfo {author} {\bibfnamefont {Zengyi}\ \bibnamefont {Du}},
  \bibinfo {author} {\bibfnamefont {Xiong}\ \bibnamefont {Yang}}, \bibinfo
  {author} {\bibfnamefont {Enyu}\ \bibnamefont {Wang}}, \bibinfo {author}
  {\bibfnamefont {Delong}\ \bibnamefont {Fang}}, \bibinfo {author}
  {\bibfnamefont {Huan}\ \bibnamefont {Yang}}, \ and\ \bibinfo {author}
  {\bibfnamefont {Hai-Hu}\ \bibnamefont {Wen}},\ }\bibfield  {title} {\enquote
  {\bibinfo {title} {Merging dirac electrons and correlation effect in the
  heterostructured bi2te3/fe1+ dte},}\ }\href@noop {} {\bibfield  {journal}
  {\bibinfo  {journal} {arXiv preprint arXiv:1509.07424}\ } (\bibinfo {year}
  {2015})}\BibitemShut {NoStop}%
\bibitem [{\citenamefont {Kunchur}\ \emph {et~al.}(2015)\citenamefont
  {Kunchur}, \citenamefont {Dean}, \citenamefont {Moghadam}, \citenamefont
  {Knight}, \citenamefont {He}, \citenamefont {Liu}, \citenamefont {Wang},
  \citenamefont {Lortz}, \citenamefont {Sou},\ and\ \citenamefont
  {Gurevich}}]{kunchur2015}%
  \BibitemOpen
  \bibfield  {author} {\bibinfo {author} {\bibfnamefont {M.~N.}\ \bibnamefont
  {Kunchur}}, \bibinfo {author} {\bibfnamefont {C.~L.}\ \bibnamefont {Dean}},
  \bibinfo {author} {\bibfnamefont {N.~Shayesteh}\ \bibnamefont {Moghadam}},
  \bibinfo {author} {\bibfnamefont {J.~M.}\ \bibnamefont {Knight}}, \bibinfo
  {author} {\bibfnamefont {Q.~L.}\ \bibnamefont {He}}, \bibinfo {author}
  {\bibfnamefont {H.}~\bibnamefont {Liu}}, \bibinfo {author} {\bibfnamefont
  {J.}~\bibnamefont {Wang}}, \bibinfo {author} {\bibfnamefont {R.}~\bibnamefont
  {Lortz}}, \bibinfo {author} {\bibfnamefont {I.~K.}\ \bibnamefont {Sou}}, \
  and\ \bibinfo {author} {\bibfnamefont {A.}~\bibnamefont {Gurevich}},\
  }\bibfield  {title} {\enquote {\bibinfo {title} {Current-induced depairing in
  the ${\mathrm{bi}}_{2}{\mathrm{te}}_{3}/\mathrm{FeTe}$ interfacial
  superconductor},}\ }\href {\doibase 10.1103/PhysRevB.92.094502} {\bibfield
  {journal} {\bibinfo  {journal} {Phys. Rev. B}\ }\textbf {\bibinfo {volume}
  {92}},\ \bibinfo {pages} {094502} (\bibinfo {year} {2015})}\BibitemShut
  {NoStop}%
\bibitem [{\citenamefont {Manna}\ \emph {et~al.}(2017)\citenamefont {Manna},
  \citenamefont {Kamlapure}, \citenamefont {Cornils}, \citenamefont
  {H{\"a}nke}, \citenamefont {Hedegaard}, \citenamefont {Bremholm},
  \citenamefont {Iversen}, \citenamefont {Hofmann}, \citenamefont {Wiebe},\
  and\ \citenamefont {Wiesendanger}}]{manna2017}%
  \BibitemOpen
  \bibfield  {author} {\bibinfo {author} {\bibfnamefont {Sujit}\ \bibnamefont
  {Manna}}, \bibinfo {author} {\bibfnamefont {Anand}\ \bibnamefont
  {Kamlapure}}, \bibinfo {author} {\bibfnamefont {Lasse}\ \bibnamefont
  {Cornils}}, \bibinfo {author} {\bibfnamefont {Torben}\ \bibnamefont
  {H{\"a}nke}}, \bibinfo {author} {\bibfnamefont {Ellen Marie~Jensen}\
  \bibnamefont {Hedegaard}}, \bibinfo {author} {\bibfnamefont {Martin}\
  \bibnamefont {Bremholm}}, \bibinfo {author} {\bibfnamefont {Bo~Brummerstedt}\
  \bibnamefont {Iversen}}, \bibinfo {author} {\bibfnamefont {Ph}~\bibnamefont
  {Hofmann}}, \bibinfo {author} {\bibfnamefont {Jens}\ \bibnamefont {Wiebe}}, \
  and\ \bibinfo {author} {\bibfnamefont {Roland}\ \bibnamefont
  {Wiesendanger}},\ }\bibfield  {title} {\enquote {\bibinfo {title}
  {Interfacial superconductivity in a bi-collinear antiferromagnetically
  ordered fete monolayer on a topological insulator},}\ }\href@noop {}
  {\bibfield  {journal} {\bibinfo  {journal} {Nature communications}\ }\textbf
  {\bibinfo {volume} {8}},\ \bibinfo {pages} {14074} (\bibinfo {year}
  {2017})}\BibitemShut {NoStop}%
\bibitem [{\citenamefont {Arnold}\ \emph {et~al.}(2018)\citenamefont {Arnold},
  \citenamefont {Warmuth}, \citenamefont {Michiardi}, \citenamefont
  {Fikáček}, \citenamefont {Bianchi}, \citenamefont {Hu}, \citenamefont
  {Mao}, \citenamefont {Miwa}, \citenamefont {Singh}, \citenamefont {Bremholm},
  \citenamefont {Wiesendanger}, \citenamefont {Honolka}, \citenamefont
  {Wehling}, \citenamefont {Wiebe},\ and\ \citenamefont
  {Hofmann}}]{fabian2018}%
  \BibitemOpen
  \bibfield  {author} {\bibinfo {author} {\bibfnamefont {Fabian}\ \bibnamefont
  {Arnold}}, \bibinfo {author} {\bibfnamefont {Jonas}\ \bibnamefont {Warmuth}},
  \bibinfo {author} {\bibfnamefont {Matteo}\ \bibnamefont {Michiardi}},
  \bibinfo {author} {\bibfnamefont {Jan}\ \bibnamefont {Fikáček}}, \bibinfo
  {author} {\bibfnamefont {Marco}\ \bibnamefont {Bianchi}}, \bibinfo {author}
  {\bibfnamefont {Jin}\ \bibnamefont {Hu}}, \bibinfo {author} {\bibfnamefont
  {Zhiqiang}\ \bibnamefont {Mao}}, \bibinfo {author} {\bibfnamefont {Jill}\
  \bibnamefont {Miwa}}, \bibinfo {author} {\bibfnamefont {Udai~Raj}\
  \bibnamefont {Singh}}, \bibinfo {author} {\bibfnamefont {Martin}\
  \bibnamefont {Bremholm}}, \bibinfo {author} {\bibfnamefont {Roland}\
  \bibnamefont {Wiesendanger}}, \bibinfo {author} {\bibfnamefont {Jan}\
  \bibnamefont {Honolka}}, \bibinfo {author} {\bibfnamefont {Tim}\ \bibnamefont
  {Wehling}}, \bibinfo {author} {\bibfnamefont {Jens}\ \bibnamefont {Wiebe}}, \
  and\ \bibinfo {author} {\bibfnamefont {Philip}\ \bibnamefont {Hofmann}},\
  }\bibfield  {title} {\enquote {\bibinfo {title} {Electronic structure of fe
  1.08 te bulk crystals and epitaxial fete thin films on bi 2 te 3},}\ }\href
  {http://stacks.iop.org/0953-8984/30/i=6/a=065502} {\bibfield  {journal}
  {\bibinfo  {journal} {Journal of Physics: Condensed Matter}\ }\textbf
  {\bibinfo {volume} {30}},\ \bibinfo {pages} {065502} (\bibinfo {year}
  {2018})}\BibitemShut {NoStop}%
\bibitem [{\citenamefont {Singh}\ \emph {et~al.}(2018)\citenamefont {Singh},
  \citenamefont {Warmuth}, \citenamefont {Kamlapure}, \citenamefont {Cornils},
  \citenamefont {Bremholm}, \citenamefont {Hofmann}, \citenamefont {Wiebe},\
  and\ \citenamefont {Wiesendanger}}]{Raj2018}%
  \BibitemOpen
  \bibfield  {author} {\bibinfo {author} {\bibfnamefont {Udai~Raj}\
  \bibnamefont {Singh}}, \bibinfo {author} {\bibfnamefont {Jonas}\ \bibnamefont
  {Warmuth}}, \bibinfo {author} {\bibfnamefont {Anand}\ \bibnamefont
  {Kamlapure}}, \bibinfo {author} {\bibfnamefont {Lasse}\ \bibnamefont
  {Cornils}}, \bibinfo {author} {\bibfnamefont {Martin}\ \bibnamefont
  {Bremholm}}, \bibinfo {author} {\bibfnamefont {Philip}\ \bibnamefont
  {Hofmann}}, \bibinfo {author} {\bibfnamefont {Jens}\ \bibnamefont {Wiebe}}, \
  and\ \bibinfo {author} {\bibfnamefont {Roland}\ \bibnamefont
  {Wiesendanger}},\ }\bibfield  {title} {\enquote {\bibinfo {title} {Enhanced
  spin-ordering temperature in ultrathin fete films grown on a topological
  insulator},}\ }\href {\doibase 10.1103/PhysRevB.97.144513} {\bibfield
  {journal} {\bibinfo  {journal} {Phys. Rev. B}\ }\textbf {\bibinfo {volume}
  {97}},\ \bibinfo {pages} {144513} (\bibinfo {year} {2018})}\BibitemShut
  {NoStop}%
\bibitem [{\citenamefont {Li}\ \emph {et~al.}(2016)\citenamefont {Li},
  \citenamefont {Yin}, \citenamefont {Wang}, \citenamefont {He}, \citenamefont
  {Ma}, \citenamefont {Xue},\ and\ \citenamefont {Chen}}]{Li2016}%
  \BibitemOpen
  \bibfield  {author} {\bibinfo {author} {\bibfnamefont {Wei}\ \bibnamefont
  {Li}}, \bibinfo {author} {\bibfnamefont {Wei-Guo}\ \bibnamefont {Yin}},
  \bibinfo {author} {\bibfnamefont {Lili}\ \bibnamefont {Wang}}, \bibinfo
  {author} {\bibfnamefont {Ke}~\bibnamefont {He}}, \bibinfo {author}
  {\bibfnamefont {Xucun}\ \bibnamefont {Ma}}, \bibinfo {author} {\bibfnamefont
  {Qi-Kun}\ \bibnamefont {Xue}}, \ and\ \bibinfo {author} {\bibfnamefont
  {Xi}~\bibnamefont {Chen}},\ }\bibfield  {title} {\enquote {\bibinfo {title}
  {Charge ordering in stoichiometric fete: Scanning tunneling microscopy and
  spectroscopy},}\ }\href {\doibase 10.1103/PhysRevB.93.041101} {\bibfield
  {journal} {\bibinfo  {journal} {Phys. Rev. B}\ }\textbf {\bibinfo {volume}
  {93}},\ \bibinfo {pages} {041101} (\bibinfo {year} {2016})}\BibitemShut
  {NoStop}%
\bibitem [{Not()}]{Note1}%
  \BibitemOpen
  \href@noop {} {}\bibinfo {note} {Hello world}\BibitemShut {NoStop}%
\bibitem [{\citenamefont {Xu}\ \emph {et~al.}(2015)\citenamefont {Xu},
  \citenamefont {Wang}, \citenamefont {Liu}, \citenamefont {Ge}, \citenamefont
  {Yang}, \citenamefont {Liu}, \citenamefont {Xu}, \citenamefont {Guan},
  \citenamefont {Gao}, \citenamefont {Qian}, \citenamefont {Liu}, \citenamefont
  {Wang}, \citenamefont {Zhang}, \citenamefont {Xue},\ and\ \citenamefont
  {Jia}}]{Xu2015}%
  \BibitemOpen
  \bibfield  {author} {\bibinfo {author} {\bibfnamefont {Jin-Peng}\
  \bibnamefont {Xu}}, \bibinfo {author} {\bibfnamefont {Mei-Xiao}\ \bibnamefont
  {Wang}}, \bibinfo {author} {\bibfnamefont {Zhi~Long}\ \bibnamefont {Liu}},
  \bibinfo {author} {\bibfnamefont {Jian-Feng}\ \bibnamefont {Ge}}, \bibinfo
  {author} {\bibfnamefont {Xiaojun}\ \bibnamefont {Yang}}, \bibinfo {author}
  {\bibfnamefont {Canhua}\ \bibnamefont {Liu}}, \bibinfo {author}
  {\bibfnamefont {Zhu~An}\ \bibnamefont {Xu}}, \bibinfo {author} {\bibfnamefont
  {Dandan}\ \bibnamefont {Guan}}, \bibinfo {author} {\bibfnamefont {Chun~Lei}\
  \bibnamefont {Gao}}, \bibinfo {author} {\bibfnamefont {Dong}\ \bibnamefont
  {Qian}}, \bibinfo {author} {\bibfnamefont {Ying}\ \bibnamefont {Liu}},
  \bibinfo {author} {\bibfnamefont {Qiang-Hua}\ \bibnamefont {Wang}}, \bibinfo
  {author} {\bibfnamefont {Fu-Chun}\ \bibnamefont {Zhang}}, \bibinfo {author}
  {\bibfnamefont {Qi-Kun}\ \bibnamefont {Xue}}, \ and\ \bibinfo {author}
  {\bibfnamefont {Jin-Feng}\ \bibnamefont {Jia}},\ }\bibfield  {title}
  {\enquote {\bibinfo {title} {Experimental detection of a majorana mode in the
  core of a magnetic vortex inside a topological insulator-superconductor
  ${\mathrm{bi}}_{2}{\mathrm{te}}_{3}/{\mathrm{nbse}}_{2}$ heterostructure},}\
  }\href {\doibase 10.1103/PhysRevLett.114.017001} {\bibfield  {journal}
  {\bibinfo  {journal} {Phys. Rev. Lett.}\ }\textbf {\bibinfo {volume} {114}},\
  \bibinfo {pages} {017001} (\bibinfo {year} {2015})}\BibitemShut {NoStop}%
\bibitem [{\citenamefont {Li}\ \emph {et~al.}(2009)\citenamefont {Li},
  \citenamefont {de~la Cruz}, \citenamefont {Huang}, \citenamefont {Chen},
  \citenamefont {Lynn}, \citenamefont {Hu}, \citenamefont {Huang},
  \citenamefont {Hsu}, \citenamefont {Yeh}, \citenamefont {Wu},\ and\
  \citenamefont {Dai}}]{Li2009}%
  \BibitemOpen
  \bibfield  {author} {\bibinfo {author} {\bibfnamefont {Shiliang}\
  \bibnamefont {Li}}, \bibinfo {author} {\bibfnamefont {Clarina}\ \bibnamefont
  {de~la Cruz}}, \bibinfo {author} {\bibfnamefont {Q.}~\bibnamefont {Huang}},
  \bibinfo {author} {\bibfnamefont {Y.}~\bibnamefont {Chen}}, \bibinfo {author}
  {\bibfnamefont {J.~W.}\ \bibnamefont {Lynn}}, \bibinfo {author}
  {\bibfnamefont {Jiangping}\ \bibnamefont {Hu}}, \bibinfo {author}
  {\bibfnamefont {Yi-Lin}\ \bibnamefont {Huang}}, \bibinfo {author}
  {\bibfnamefont {Fong-Chi}\ \bibnamefont {Hsu}}, \bibinfo {author}
  {\bibfnamefont {Kuo-Wei}\ \bibnamefont {Yeh}}, \bibinfo {author}
  {\bibfnamefont {Maw-Kuen}\ \bibnamefont {Wu}}, \ and\ \bibinfo {author}
  {\bibfnamefont {Pengcheng}\ \bibnamefont {Dai}},\ }\bibfield  {title}
  {\enquote {\bibinfo {title} {First-order magnetic and structural phase
  transitions in
  ${\text{fe}}_{1+y}{\text{se}}_{x}{\text{te}}_{1\ensuremath{-}x}$},}\ }\href
  {\doibase 10.1103/PhysRevB.79.054503} {\bibfield  {journal} {\bibinfo
  {journal} {Phys. Rev. B}\ }\textbf {\bibinfo {volume} {79}},\ \bibinfo
  {pages} {054503} (\bibinfo {year} {2009})}\BibitemShut {NoStop}%
\bibitem [{\citenamefont {R\"o\ss{}ler}\ \emph {et~al.}(2011)\citenamefont
  {R\"o\ss{}ler}, \citenamefont {Cherian}, \citenamefont {Lorenz},
  \citenamefont {Doerr}, \citenamefont {Koz}, \citenamefont {Curfs},
  \citenamefont {Prots}, \citenamefont {R\"o\ss{}ler}, \citenamefont {Schwarz},
  \citenamefont {Elizabeth},\ and\ \citenamefont {Wirth}}]{Roesler2011}%
  \BibitemOpen
  \bibfield  {author} {\bibinfo {author} {\bibfnamefont {S.}~\bibnamefont
  {R\"o\ss{}ler}}, \bibinfo {author} {\bibfnamefont {Dona}\ \bibnamefont
  {Cherian}}, \bibinfo {author} {\bibfnamefont {W.}~\bibnamefont {Lorenz}},
  \bibinfo {author} {\bibfnamefont {M.}~\bibnamefont {Doerr}}, \bibinfo
  {author} {\bibfnamefont {C.}~\bibnamefont {Koz}}, \bibinfo {author}
  {\bibfnamefont {C.}~\bibnamefont {Curfs}}, \bibinfo {author} {\bibfnamefont
  {Yu.}\ \bibnamefont {Prots}}, \bibinfo {author} {\bibfnamefont {U.~K.}\
  \bibnamefont {R\"o\ss{}ler}}, \bibinfo {author} {\bibfnamefont
  {U.}~\bibnamefont {Schwarz}}, \bibinfo {author} {\bibfnamefont {Suja}\
  \bibnamefont {Elizabeth}}, \ and\ \bibinfo {author} {\bibfnamefont
  {S.}~\bibnamefont {Wirth}},\ }\bibfield  {title} {\enquote {\bibinfo {title}
  {First-order structural transition in the magnetically ordered phase of
  fe${}_{1.13}$te},}\ }\href {\doibase 10.1103/PhysRevB.84.174506} {\bibfield
  {journal} {\bibinfo  {journal} {Phys. Rev. B}\ }\textbf {\bibinfo {volume}
  {84}},\ \bibinfo {pages} {174506} (\bibinfo {year} {2011})}\BibitemShut
  {NoStop}%
\bibitem [{\citenamefont {Rodriguez}\ \emph {et~al.}(2011)\citenamefont
  {Rodriguez}, \citenamefont {Stock}, \citenamefont {Zajdel}, \citenamefont
  {Krycka}, \citenamefont {Majkrzak}, \citenamefont {Zavalij},\ and\
  \citenamefont {Green}}]{Rodriguez2011}%
  \BibitemOpen
  \bibfield  {author} {\bibinfo {author} {\bibfnamefont {E.~E.}\ \bibnamefont
  {Rodriguez}}, \bibinfo {author} {\bibfnamefont {C.}~\bibnamefont {Stock}},
  \bibinfo {author} {\bibfnamefont {P.}~\bibnamefont {Zajdel}}, \bibinfo
  {author} {\bibfnamefont {K.~L.}\ \bibnamefont {Krycka}}, \bibinfo {author}
  {\bibfnamefont {C.~F.}\ \bibnamefont {Majkrzak}}, \bibinfo {author}
  {\bibfnamefont {P.}~\bibnamefont {Zavalij}}, \ and\ \bibinfo {author}
  {\bibfnamefont {M.~A.}\ \bibnamefont {Green}},\ }\bibfield  {title} {\enquote
  {\bibinfo {title} {Magnetic-crystallographic phase diagram of the
  superconducting parent compound fe${}_{1+x}$te},}\ }\href {\doibase
  10.1103/PhysRevB.84.064403} {\bibfield  {journal} {\bibinfo  {journal} {Phys.
  Rev. B}\ }\textbf {\bibinfo {volume} {84}},\ \bibinfo {pages} {064403}
  (\bibinfo {year} {2011})}\BibitemShut {NoStop}%
\bibitem [{\citenamefont {Mizuguchi}\ \emph {et~al.}(2012)\citenamefont
  {Mizuguchi}, \citenamefont {Hamada}, \citenamefont {Goto}, \citenamefont
  {Takatsu}, \citenamefont {Kadowaki},\ and\ \citenamefont
  {Miura}}]{Mizuguchi2012}%
  \BibitemOpen
  \bibfield  {author} {\bibinfo {author} {\bibfnamefont {Yoshikazu}\
  \bibnamefont {Mizuguchi}}, \bibinfo {author} {\bibfnamefont {Kentaro}\
  \bibnamefont {Hamada}}, \bibinfo {author} {\bibfnamefont {Kazuki}\
  \bibnamefont {Goto}}, \bibinfo {author} {\bibfnamefont {Hiroshi}\
  \bibnamefont {Takatsu}}, \bibinfo {author} {\bibfnamefont {Hiroaki}\
  \bibnamefont {Kadowaki}}, \ and\ \bibinfo {author} {\bibfnamefont {Osuke}\
  \bibnamefont {Miura}},\ }\bibfield  {title} {\enquote {\bibinfo {title}
  {Evolution of two-step structural phase transition in fe1+dte detected by
  low-temperature x-ray diffraction},}\ }\href {\doibase
  https://doi.org/10.1016/j.ssc.2012.03.022} {\bibfield  {journal} {\bibinfo
  {journal} {Solid State Communications}\ }\textbf {\bibinfo {volume} {152}},\
  \bibinfo {pages} {1047 -- 1051} (\bibinfo {year} {2012})}\BibitemShut
  {NoStop}%
\bibitem [{\citenamefont {Zaliznyak}\ \emph {et~al.}(2012)\citenamefont
  {Zaliznyak}, \citenamefont {Xu}, \citenamefont {Wen}, \citenamefont
  {Tranquada}, \citenamefont {Gu}, \citenamefont {Solovyov}, \citenamefont
  {Glazkov}, \citenamefont {Zheludev}, \citenamefont {Garlea},\ and\
  \citenamefont {Stone}}]{Zaliznyak2012}%
  \BibitemOpen
  \bibfield  {author} {\bibinfo {author} {\bibfnamefont {I.~A.}\ \bibnamefont
  {Zaliznyak}}, \bibinfo {author} {\bibfnamefont {Z.~J.}\ \bibnamefont {Xu}},
  \bibinfo {author} {\bibfnamefont {J.~S.}\ \bibnamefont {Wen}}, \bibinfo
  {author} {\bibfnamefont {J.~M.}\ \bibnamefont {Tranquada}}, \bibinfo {author}
  {\bibfnamefont {G.~D.}\ \bibnamefont {Gu}}, \bibinfo {author} {\bibfnamefont
  {V.}~\bibnamefont {Solovyov}}, \bibinfo {author} {\bibfnamefont {V.~N.}\
  \bibnamefont {Glazkov}}, \bibinfo {author} {\bibfnamefont {A.~I.}\
  \bibnamefont {Zheludev}}, \bibinfo {author} {\bibfnamefont {V.~O.}\
  \bibnamefont {Garlea}}, \ and\ \bibinfo {author} {\bibfnamefont {M.~B.}\
  \bibnamefont {Stone}},\ }\bibfield  {title} {\enquote {\bibinfo {title}
  {Continuous magnetic and structural phase transitions in fe${}_{1+y}$te},}\
  }\href {\doibase 10.1103/PhysRevB.85.085105} {\bibfield  {journal} {\bibinfo
  {journal} {Phys. Rev. B}\ }\textbf {\bibinfo {volume} {85}},\ \bibinfo
  {pages} {085105} (\bibinfo {year} {2012})}\BibitemShut {NoStop}%
\bibitem [{\citenamefont {Koz}\ \emph {et~al.}(2013)\citenamefont {Koz},
  \citenamefont {R\"o\ss{}ler}, \citenamefont {Tsirlin}, \citenamefont
  {Wirth},\ and\ \citenamefont {Schwarz}}]{Koz2013}%
  \BibitemOpen
  \bibfield  {author} {\bibinfo {author} {\bibfnamefont {Cevriye}\ \bibnamefont
  {Koz}}, \bibinfo {author} {\bibfnamefont {Sahana}\ \bibnamefont
  {R\"o\ss{}ler}}, \bibinfo {author} {\bibfnamefont {Alexander~A.}\
  \bibnamefont {Tsirlin}}, \bibinfo {author} {\bibfnamefont {Steffen}\
  \bibnamefont {Wirth}}, \ and\ \bibinfo {author} {\bibfnamefont {Ulrich}\
  \bibnamefont {Schwarz}},\ }\bibfield  {title} {\enquote {\bibinfo {title}
  {Low-temperature phase diagram of fe${}_{1+y}$te studied using x-ray
  diffraction},}\ }\href {\doibase 10.1103/PhysRevB.88.094509} {\bibfield
  {journal} {\bibinfo  {journal} {Phys. Rev. B}\ }\textbf {\bibinfo {volume}
  {88}},\ \bibinfo {pages} {094509} (\bibinfo {year} {2013})}\BibitemShut
  {NoStop}%
\bibitem [{\citenamefont {Si}\ \emph {et~al.}(2010)\citenamefont {Si},
  \citenamefont {Jie}, \citenamefont {Wu}, \citenamefont {Zhou}, \citenamefont
  {Gu}, \citenamefont {Johnson},\ and\ \citenamefont {Li}}]{Si2010}%
  \BibitemOpen
  \bibfield  {author} {\bibinfo {author} {\bibfnamefont {Weidong}\ \bibnamefont
  {Si}}, \bibinfo {author} {\bibfnamefont {Qing}\ \bibnamefont {Jie}}, \bibinfo
  {author} {\bibfnamefont {Lijun}\ \bibnamefont {Wu}}, \bibinfo {author}
  {\bibfnamefont {Juan}\ \bibnamefont {Zhou}}, \bibinfo {author} {\bibfnamefont
  {Genda}\ \bibnamefont {Gu}}, \bibinfo {author} {\bibfnamefont {P.~D.}\
  \bibnamefont {Johnson}}, \ and\ \bibinfo {author} {\bibfnamefont {Qiang}\
  \bibnamefont {Li}},\ }\bibfield  {title} {\enquote {\bibinfo {title}
  {Superconductivity in epitaxial thin films of
  ${\text{fe}}_{1.08}\text{Te}:{\text{o}}_{x}$},}\ }\href {\doibase
  10.1103/PhysRevB.81.092506} {\bibfield  {journal} {\bibinfo  {journal} {Phys.
  Rev. B}\ }\textbf {\bibinfo {volume} {81}},\ \bibinfo {pages} {092506}
  (\bibinfo {year} {2010})}\BibitemShut {NoStop}%
\bibitem [{\citenamefont {Schlaf}()}]{Schlaf2013}%
  \BibitemOpen
  \bibfield  {author} {\bibinfo {author} {\bibfnamefont {Rudy}\ \bibnamefont
  {Schlaf}},\ }\bibfield  {title} {\enquote {\bibinfo {title} {Calibration of
  photoemission spectra and work function determination},}\ }\href@noop {} {\
  }\Eprint {http://arxiv.org/abs/http://rsl.eng.usf.edu/Documents/Tutorials/
  PEScalibration.pdf} {http://rsl.eng.usf.edu/Documents/Tutorials/
  PEScalibration.pdf} \BibitemShut {NoStop}%
\bibitem [{\citenamefont {Hor}\ \emph {et~al.}(2011)\citenamefont {Hor},
  \citenamefont {Checkelsky}, \citenamefont {Qu}, \citenamefont {Ong},\ and\
  \citenamefont {Cava}}]{Hor2010}%
  \BibitemOpen
  \bibfield  {author} {\bibinfo {author} {\bibfnamefont {Y.S.}\ \bibnamefont
  {Hor}}, \bibinfo {author} {\bibfnamefont {J.G.}\ \bibnamefont {Checkelsky}},
  \bibinfo {author} {\bibfnamefont {D.}~\bibnamefont {Qu}}, \bibinfo {author}
  {\bibfnamefont {N.P.}\ \bibnamefont {Ong}}, \ and\ \bibinfo {author}
  {\bibfnamefont {R.J.}\ \bibnamefont {Cava}},\ }\bibfield  {title} {\enquote
  {\bibinfo {title} {Superconductivity and non-metallicity induced by doping
  the topological insulators bi2se3 and bi2te3},}\ }\href {\doibase
  https://doi.org/10.1016/j.jpcs.2010.10.027} {\bibfield  {journal} {\bibinfo
  {journal} {Journal of Physics and Chemistry of Solids}\ }\textbf {\bibinfo
  {volume} {72}},\ \bibinfo {pages} {572 -- 576} (\bibinfo {year} {2011})},\
  \bibinfo {note} {spectroscopies in Novel Superconductors 2010}\BibitemShut
  {NoStop}%
\bibitem [{\citenamefont {Fang}\ \emph {et~al.}(2008)\citenamefont {Fang},
  \citenamefont {Pham}, \citenamefont {Qian}, \citenamefont {Liu},
  \citenamefont {Vehstedt}, \citenamefont {Liu}, \citenamefont {Spinu},\ and\
  \citenamefont {Mao}}]{fang2008}%
  \BibitemOpen
  \bibfield  {author} {\bibinfo {author} {\bibfnamefont {M.~H.}\ \bibnamefont
  {Fang}}, \bibinfo {author} {\bibfnamefont {H.~M.}\ \bibnamefont {Pham}},
  \bibinfo {author} {\bibfnamefont {B.}~\bibnamefont {Qian}}, \bibinfo {author}
  {\bibfnamefont {T.~J.}\ \bibnamefont {Liu}}, \bibinfo {author} {\bibfnamefont
  {E.~K.}\ \bibnamefont {Vehstedt}}, \bibinfo {author} {\bibfnamefont
  {Y.}~\bibnamefont {Liu}}, \bibinfo {author} {\bibfnamefont {L.}~\bibnamefont
  {Spinu}}, \ and\ \bibinfo {author} {\bibfnamefont {Z.~Q.}\ \bibnamefont
  {Mao}},\ }\bibfield  {title} {\enquote {\bibinfo {title} {Superconductivity
  close to magnetic instability in
  $\text{Fe}{({\text{Se}}_{1\ensuremath{-}x}{\text{Te}}_{x})}_{0.82}$},}\
  }\href {\doibase 10.1103/PhysRevB.78.224503} {\bibfield  {journal} {\bibinfo
  {journal} {Phys. Rev. B}\ }\textbf {\bibinfo {volume} {78}},\ \bibinfo
  {pages} {224503} (\bibinfo {year} {2008})}\BibitemShut {NoStop}%
\end{thebibliography}%
\end{document}